\documentclass{aip-cp}

\usepackage[numbers]{natbib}
\usepackage{rotating}
\usepackage{graphicx}

\usepackage{combelow}
\usepackage{bm}
\usepackage{slashed}
\usepackage{braket}
\usepackage{color}

\def\eqref#1{(\ref{#1})}

\begin{document}

\title{Quantum Corrections in Thermal States of Fermions on Anti-de Sitter Space-time}

\author[aff1]{Victor E. Ambru\cb{s}\corref{cor1}}

\author[aff2]{Elizabeth Winstanley\corref{cor2}}

\affil[aff1]{Department of Physics, West University of Timi\cb{s}oara,
	Bd.~Vasile P\^arvan 4, Timi\cb{s}oara 300223, Romania}
\affil[aff2]{Consortium for Fundamental Physics, School of Mathematics and Statistics, University of Sheffield,
	Hicks Building, Hounsfield Road, Sheffield. S3 7RH United Kingdom}
\corresp[cor1]{Victor.Ambrus@e-uvt.ro}
\corresp[cor2]{Corresponding author: E.Winstanley@sheffield.ac.uk}

\maketitle

\begin{abstract}
We study the energy density and pressure of a relativistic thermal gas of massless fermions on four-dimensional Minkowski and anti-de Sitter space-times using relativistic kinetic theory.
The corresponding quantum field theory quantities are given by components of the renormalized expectation value of the stress-energy tensor operator acting on a thermal state.
On Minkowski space-time, the renormalized vacuum expectation value of the stress-energy tensor is by definition
zero, while on anti-de Sitter space-time the vacuum contribution to this expectation value is in general
nonzero. We compare the properties of the vacuum and thermal expectation values of
the energy density and pressure for massless fermions and discuss the circumstances in which the thermal contribution dominates over the vacuum one.
\end{abstract}

\section{INTRODUCTION}
\label{sec:intro}

There has been an explosion of interest in physics on anti-de Sitter (adS) space-time in the twenty years following the formulation of the adS/CFT (conformal field theory) correspondence (see \cite{Aharony:1999ti} for a review).
According to this correspondence, quantum gravity on certain asymptotically adS space-times is dual to a conformal field theory living on the boundary of the particular space-time.
As a semi-classical approximation to full quantum gravity, one may consider quantum fields on a fixed asymptotically adS space-time.
In this paper we consider the simplest asymptotically adS space-time, namely pure adS space-time itself.

In previous work \cite{Ambrus:2015mfa, Ambrus:2014fka}, we have studied the quantum vacuum expectation value
(v.e.v.) of the stress-energy tensor (SET) for fermions on this background.  We found that v.e.v.s for
fermions on adS are nonzero, in contrast to the situation in Minkowski space-time, where the v.e.v. of the
SET vanishes (by definition).
In \cite{Ambrus:2017cow},
the SET of a thermal configuration of fermions on adS is considered in two complementary frameworks:
relativistic kinetic theory (RKT) and quantum field theory (QFT) in curved space-time.
By considering the difference between the thermal expectation value (t.e.v.) and the v.e.v.
of the SET, the QFT and the RKT results for the energy density and pressure are compared.
The conclusion in \cite{Ambrus:2017cow} is that at low temperatures,  quantum corrections
are significant.
The situation in Minkowski space is markedly different.
We will show that the t.e.v.~of the SET is exactly equal to the RKT prediction in this case, in other words,
quantum corrections play no role on flat space-time.

In \cite{Ambrus:2017cow} we focussed on the difference between the t.e.v.~and the v.e.v.~of the SET for fermions on adS.
The purpose of this paper is to consider the full t.e.v.~with vacuum contributions included.
We will identify the regime where thermal contributions to the t.e.v.~of the SET dominate over the
underlying vacuum contributions. This turns out to be a nontrivial question, since the v.e.v. is constant throughout adS space-time, while the
t.e.v.~of the SET is position-dependent.
Our analysis is mainly concerned with massless fermions since for massive fermions the v.e.v.~of the SET depends on both the method of renormalization and an arbitrary mass renormalization scale \cite{Ambrus:2015mfa}.

Throughout this paper, we use Planck units with $k_B = c = G = \hbar = 1$ and we employ the
$(-, +, +, +)$ metric signature.

\section{THERMAL STATES OF FERMIONS ON MINKOWSKI SPACE-TIME}
\label{sec:mink}

In this section we briefly review the corresponding calculations in both RKT and QFT on Minkowski space-time.  For ease of comparison with our later results on adS, we denote Minkowski space-time indices
 using hats, e.g.~${\hat {\alpha }}$.

\subsection{Relativistic kinetic theory}
\label{sec:MinkRKT}

In RKT, a thermal gas of fermion particles of mass $m$ and momentum $p^{{\hat {\alpha }}}$
at inverse temperature $\beta $ is described by the Fermi-Dirac distribution function
\cite{Ambrus:2016ocv, Ambrus2015, Florkowski:2014sda} with $Z=4$ degrees of freedom per particle
\begin{equation}
f_{\beta } = \frac{Z}{\left( 2\pi \right) ^{3} \left( e^{\beta p^{\hat{0}} } + 1 \right)}
=
\frac {1}{2 \pi^{3}}
\sum_{j = 1}^\infty (-1)^{j-1}
e^{-j\beta p^{\hat{0}}},
\label{eq:feq}
\end{equation}
where we have assumed that the chemical potential vanishes and consider a state where the fluid is at rest.
The thermal SET is given by integrating the distribution function \eqref{eq:feq} with respect to the particle momentum
\cite{Ambrus:2016ocv, Ambrus2015}:
\begin{equation}
T^{\hat{\alpha}\hat{\sigma}} = \int \frac{d^3p}{p^{\hat{0}}} f_{\beta }\, p^{\hat{\alpha }} p^{\hat {\sigma}}
= {\mathrm {Diag}} \left\{ E^{{\mathrm {M}}}(\beta ) , P^{{\mathrm {M}}}(\beta ), P^{{\mathrm {M}}}(\beta ),
P^{{\mathrm {M}}}(\beta ) \right\} .
\label{eq:fluid}
\end{equation}
The SET \eqref{eq:fluid} has perfect fluid form, and the energy density
$E^{{\mathrm {M}}}(\beta )$ and pressure $P^{{\mathrm {M}}}(\beta )$ are given
by
\begin{equation}
E^{{\mathrm{M}}}(\beta )- 3P^{\mathrm  {M}}(\beta ) = \frac{2m^3}{\pi^2 \beta}
\sum_{j = 1}^\infty \frac{(-1)^{j-1}}{j} K_1\left(m j\beta\right),
\qquad
P^{{\mathrm {M}}}(\beta )  = \frac{2 m^2}{\pi^2 \beta^2}
\sum_{j = 1}^\infty \frac{(-1)^{j-1}}{j^2} K_2\left(m j\beta\right),
\label{eq:tevM}
\end{equation}
where $K_{n}(z)$ is a modified Bessel function.
Using the asymptotic properties of modified Bessel functions, the energy density and pressure in the massless limit $m\rightarrow 0$ can be found
to be
\begin{equation}
\left. E^{{\mathrm {M}}}(\beta ) \right\rfloor _{m=0} =
\left. 3 P^{{\mathrm {M}}}(\beta ) \right\rfloor _{m=0}  =
\frac{7\pi ^{2}}{60 \beta^4} .
\label{eq:mzeroM}
\end{equation}
The energy density $E^{{\mathrm{M}}}(\beta ) $ and pressure $P^{{\mathrm{M}}}(\beta )$
are space-time constants, depending only on $m$ and $\beta $.
For zero temperature
($\beta \rightarrow \infty$), both the energy density and pressure vanish
and the vacuum state is achieved.

\subsection{Quantum field theory}
\label{sec:MinkQFT}

In QFT, in order to compute expectation values, we require the Minkowski space-time thermal Feynman Green's function $S^{{\mathrm {M}}}_{\beta }(x,x')$, which satisfies the inhomogeneous Dirac equation for fermions of mass $m$
\begin{equation}
\left( i \slashed{\partial } - m \right) S^{\mathrm {M}}_{\beta }(x,x') =  \delta^4(x - x') ,
\label{eq:sf_eq}
\end{equation}
where $\slashed{\partial } = \gamma ^{\hat {\alpha }}\partial _{\hat {\alpha }}$ and
$\gamma ^{\hat {\alpha }}$ are the usual Dirac matrices, given in \cite{Ambrus:2017cow}.
The Minkowski vacuum Feynman Green's function $S^{{\mathrm {M}}}_{\mathrm {vac}}(x,x')$
takes the form \cite{Muck:1999mh}
(where $H^{(2)}_{1}$ is a Hankel function of the second kind)
\begin{equation}
iS^{{\mathrm {M}}}_{\mathrm {vac}} (x,x') = \left(i\slashed{\partial} + m \right)
\frac{{\mathcal {A}}_{{\mathrm{M}}}}{m},
\qquad
{\mathcal {A}}_{\mathrm{M}} = \frac{im^{2}}{8\pi s_{\mathrm {M}}}
H^{(2)}_{1}(ms_{\mathrm {M}}).
\label{eq:SFvacM}
\end{equation}
The geodetic interval $s_{\mathrm {M}}(x,x')$ between the space-time points $x=(t,\bm{x})$ and $x'=(t',{\bm{x'}})$ is given, for Minkowski space-time, by
$s_{\mathrm {M}}^{2} = -\left( x^{\mu }-x^{\mu'} \right)\left(x_{\mu } - x_{\mu'}\right)$.
The thermal Feynman Green's function $S^{{\mathrm {M}}}_{\beta }(x,x')$ is constructed from the vacuum Feynman Green's function
(\ref{eq:SFvacM}) as follows \cite{Birrell:1983}:
\begin{equation}
S^{\mathrm {M}}_\beta(x, x') = \sum_{j=-\infty }^{\infty } (-1)^j S^{\mathrm {M}}_{\mathrm {vac}}(t + ij\beta, \bm{x}; t', \bm{x}') .
\label{eq:SFbetaM}
\end{equation}
The t.e.v.~of the SET is computed from $S^{\mathrm{M}}_\beta(x,x')$ using  the following expression \cite{Ambrus:2015mfa}:
\begin{equation}
\braket{T_{\hat{\alpha}\hat{\rho}}}^{\mathrm {M}}_\beta = \frac{i}{2} \lim_{x'\rightarrow x} {\mathrm {tr}}\left\{
\left[\gamma_{(\hat{\alpha}} \partial _{\hat{\rho})} iS^{\mathrm {M}}_{\beta }(x, x') -
\partial _{\hat{\rho}'} [iS^{\mathrm {M}}_{\beta }(x,x')]
\gamma_{(\hat{\alpha}} \delta  _{\hat {\rho })}^{{\hat {\rho}}'}
\right]  \right\}.
\label{eq:ev_set}
\end{equation}
Taking the limit $x'\rightarrow x$ results in a quantity which is infinite.
On Minkowski space-time, this is renormalized by subtracting $S^{\mathrm{M}}_{\mathrm {vac}}(x,x')$ (\ref{eq:SFvacM})
(corresponding to $j=0$) from (\ref{eq:SFbetaM}), and then \eqref{eq:ev_set} becomes:
\begin{equation}
\braket{:T_{\hat{\alpha}\hat{\rho}}:}_{\beta }^{\mathrm {M}} \equiv
\braket{T_{\hat{\alpha}\hat{\rho}}}_{\beta }^{\mathrm {M}} -
\braket{0| T_{\hat{\alpha}\hat{\rho}} |0} =
-\frac{2}{m}
\lim_{x'\rightarrow x} \sum_{j \neq 0} (-1)^j \left\{
 2\eta_{\hat{\alpha}\hat{\rho}} \frac{{\mathcal {A}}_{\mathrm {M}}'}{s_{\mathrm {M}}} +
 [n_{(\hat{\alpha}} - n_{(\hat{\alpha}'}] n_{\hat{\rho})} \left(\frac{{\mathcal {A}}_{\mathrm {M}}'}{s_{\mathrm {M}}} - {\mathcal {A}} _{\mathrm{M}}''\right)\right\},
 \label{eq:tev_set}
\end{equation}
where $\eta _{{\hat {\alpha }}{\hat {\rho}}}$ is the Minkowski metric and the prime denotes a derivative with respect to $s_{\mathrm {M}}$,
while $n_{\hat{\alpha}} = \partial_{\hat{\alpha}} s_{\mathrm {M}}$ and
$n_{\hat{\alpha}'} = \partial_{\hat{\alpha}'} s_{\mathrm {M}}$ are the tangents to the geodesic connecting $x$ and
$x'$ at the points $x$ and $x'$ respectively.
Taking the limit $t'\rightarrow t$, ${\bm {x'}} \rightarrow {\bm {x}}$, the vacuum Feynman Green's function in (\ref{eq:SFbetaM}) depends on the geodetic interval $s_{M}$ between the points $x=(t,{\bm {x}})$ and $x'=(t+ij\beta, {\bm {x}})$, which reduces to
$s_j = e^{-i\pi / 2} |j| \beta$, and then  $H^{(2)}_1(m s_j) = -2K_1(m |j|\beta)/\pi $.
The resulting t.e.v.~of the SET has the perfect fluid form (\ref{eq:fluid})
with the energy density and pressure being identical to those calculated in RKT (\ref{eq:tevM}).
We therefore conclude that there are no quantum corrections to the thermal SET on flat space-time.

\section{THERMAL STATES OF FERMIONS ON ANTI-DE SITTER SPACE-TIME}
\label{sec:adS}

In this section we examine the transition from the Minkowski case, when the RKT and
the QFT results coincide, to adS space-time,
where it is shown in \cite{Ambrus:2017cow} that quantum corrections can become significant.

AdS is a maximally-symmetric space-time, whose metric, in four dimensions, takes the form:
\begin{equation}
ds^{2} = \frac {1}{(\cos \omega r)^2 } \left[ -dt^{2} + dr^{2}
+ \frac {\sin ^{2}\omega r}{\omega ^{2}} \left( d\theta ^{2} +
\sin ^{2} \theta \, d\varphi ^{2} \right) \right] ,
\label{eq:metric}
\end{equation}
where $\omega $ is the inverse radius of curvature.
The time coordinate $t$ has a finite range for adS, but this leads to closed time-like curves.
To avoid these pathologies, we consider the covering space of adS, where
$t$ is allowed to take any real value.
The radial coordinate $r\in [0, \pi /2\omega ]$, with $r=\pi /2\omega $ being the time-like boundary
of adS, while
$\theta$ and $\varphi$ are the usual
spherical polar coordinates.
The boundary at $r=\pi/ 2\omega $ is an infinite proper distance from the origin $r=0$ of the space-time and hence cannot be approached by time-like geodesics in finite proper time.
Nonetheless, it can be reached in finite affine time by a null geodesic.

To aid comparison with the Minkowski space-time results obtained in the previous section, we write
the components of the SET in both RKT and QFT relative to the following Cartesian-gauge
tetrad \cite{Cotaescu:2007xv}
\begin{equation}
e_{\hat{0}}  =   \cos \omega r \, \partial_t, \qquad
e_{\hat{i}} = \cos \omega r \left[ \frac{\omega r}{\sin \omega r} \left(
\delta_{ij} - \frac{x^ix^j}{r^2}\right) + \frac{x^ix^j}{r^2}\right]
\partial_j.
\label{eq:frame}
\end{equation}

\subsection{Relativistic kinetic theory}
\label{sec:kinetic}

On a curved space-time, the inverse temperature $\beta $ in the thermal distribution function $f_{\beta }$ (\ref{eq:feq}) is replaced by the local inverse temperature ${\tilde {\beta }}$. On adS this takes the form
\begin{equation}
{\tilde {\beta }} = \beta\sqrt{-g_{tt}} =
\frac{\beta}{\cos\omega r},
\label{eq:tolman}
\end{equation}
where $\beta =  {\tilde {\beta }}(r = 0)$ is the local inverse temperature at the coordinate origin.
The calculation of the energy density $E^{{\rm {adS}}}_{\mathrm {RKT}}(\beta )$ and pressure
$P^{{\rm {adS}}}_{\mathrm {RKT}}(\beta )$ of a thermal gas of fermions on adS follows that in
Minkowski space-time, but with $\beta $ in (\ref{eq:tevM}) replaced by ${\tilde {\beta}}$.
We therefore find \cite{Ambrus:2017cow}
\begin{equation}
E^{{\mathrm {adS}}}_{\mathrm {RKT}}(\beta )- 3P^{{\mathrm {adS}}}_{\mathrm {RKT}}(\beta) = \frac{2m^3\cos \omega r}{\pi^2 \beta}
\sum_{j = 1}^\infty \frac{(-1)^{j-1}}{j} K_1\left( \frac{m j\beta}{\cos \omega r}\right),
\qquad
P^{{\mathrm {adS}}}_{\mathrm {RKT}}(\beta )  = \frac{2 m^2\cos ^{2}\omega r}{\pi^2 \beta^2}
\sum_{j = 1}^\infty \frac{(-1)^{j-1}}{j^2} K_2\left(\frac{m j\beta}{\cos \omega r}\right).
\label{eq:tevadS}
\end{equation}
The massless limit $m\rightarrow 0$ is now \cite{Ambrus:2017cow}
\begin{equation}
\left. E^{{\mathrm {adS}}}_{\mathrm{RKT}}(\beta ) \right\rfloor _{m=0} = \left. 3 P^{{\mathrm {adS}}}_{\mathrm {RKT}}(\beta ) \right\rfloor _{m=0}  = \frac{7\pi ^{2}}{60 \beta^4} \cos ^{4} \omega r.
\label{eq:mzeroadS}
\end{equation}

\pagebreak

Since on adS space-time the local inverse temperature ${\tilde {\beta }}$ (\ref{eq:tolman}) is not constant,
$E^{{\mathrm {adS}}}_{\mathrm{RKT}}(\beta )$ and
$P^{{\mathrm {adS}}}_{\mathrm{RKT}}(\beta )$ depend on $r$.
Both $E^{{\mathrm {adS}}}_{\mathrm{RKT}}(\beta )$ and
$P^{{\mathrm {adS}}}_{\mathrm {RKT}}(\beta )$ are zero
as the space-time boundary is approached ($\omega r \rightarrow \pi/2$), since
$\cos \omega r \rightarrow 0 $, ${\tilde {\beta }} \rightarrow \infty$ (\ref{eq:tolman}) and
the local temperature vanishes on the boundary.
Setting $\omega r =0$  in (\ref{eq:tevadS}, \ref{eq:mzeroadS}) gives the
Minkowski space-time results (\ref{eq:tevM}, \ref{eq:mzeroM}) respectively.
Hence, at the origin, the curvature of adS space-time does not affect the classical energy density or pressure.

\subsection{Quantum field theory}
\label{sec:QFT}

The thermal Feynman Green's function $S^{{\mathrm{adS}}}_{\beta }(x,x')$ on adS satisfies the curved-space analogue of (\ref{eq:sf_eq}):
\begin{equation}
\left( i \slashed{D} - m \right) S^{\mathrm{adS}}_{\beta }(x,x') = \left(-g \right) ^{-\frac {1}{2}} \delta^4(x - x') ,
\label{eq:sf_eq_adS}
\end{equation}
where $g$ is the determinant of the metric (\ref{eq:metric}) and $\slashed{D}$ is the spinor covariant derivative (see \cite{Ambrus:2017cow} for details).
Since adS is maximally symmetric,  the vacuum Feynman Green's function
$S^{\mathrm {adS}}_{\mathrm{vac}}(x,x')$ can be written in closed form \cite{Muck:1999mh}
\begin{equation}
i S^{\mathrm {adS}}_{\mathrm {vac}}(x,x') = ({\mathcal {A}}_{{\mathrm {adS}}} + {\mathcal {B}}_{{\mathrm {adS}}} \slashed{n}) \Lambda(x,x'),
\label{eq:sf_muck}
\end{equation}
where the scalar functions ${\mathcal {A}}_{\mathrm {adS}}$ and ${\mathcal {B}}_{\mathrm {adS}}$ depend only on the  adS  geodetic interval
$s_{\mathrm {adS}}(x,x')$, which is given by \cite{Allen:1985wd}
\begin{equation}
\cos(\omega s_{\mathrm {adS}}) = \frac{\cos\omega \Delta t}{\cos\omega r \cos\omega r'} - \cos\gamma \tan\omega r \tan\omega r',
\label{eq:geodetic}
\end{equation}
with $\Delta t = t-t'$ and $\cos\gamma = \bm{x} \cdot \bm{x'} / rr'$.
Closed form expressions can be found for
${\mathcal {A}}_{\mathrm {adS}}$ and ${\mathcal {B}}_{\mathrm {adS}}$
\cite{Ambrus:2015mfa,Ambrus:2014fka,Ambrus:2017cow,Muck:1999mh},
as well as for the bispinor of parallel transport $\Lambda(x,x')$
\cite{Ambrus:2014fka,Ambrus:2017cow}.

The construction of $S^{\mathrm {adS}}_{\beta} (x,x')$ from
$S^{\mathrm {adS}}_{\mathrm {vac}} (x,x')$ proceeds as on Minkowski space-time,
using (\ref{eq:SFbetaM}) but with $S^{\mathrm {M}}_{\mathrm {vac}}$ replaced by $S^{\mathrm {adS}}_{\mathrm {vac}}$ (\ref{eq:sf_muck}).
The t.e.v.~of the SET is then computed from $S^{\mathrm {adS}}_{\beta} (x,x')$ using the curved
space-time equivalent of (\ref{eq:ev_set}), which is
\begin{equation}
\braket{T_{\hat{\alpha}{\hat {\rho}}}}_{\beta }^{\mathrm {adS}} = \frac{i}{2} \lim_{x'\rightarrow x} {\mathrm {tr}}\left\{
\left[\gamma_{({\hat{\alpha}}} D_{\hat{\rho})} iS^{\mathrm{adS}}_{\beta }(x, x') -
D_{\hat{\rho'}} [iS^{\mathrm{adS}}_{\beta }(x,x')]
\gamma_{\hat{\alpha}'} g^{\hat{\alpha}'}{}_{({\hat{\alpha}}}
g^{{\hat{\rho}}'}{}_{\hat{\rho})}\right] \Lambda(x',x)\right\},
\label{eq:ev_set_adS}
\end{equation}
where $g_{\hat{\alpha}\hat{\rho}'}(x,x')$ is the bivector of parallel transport.
The t.e.v.~obtained from (\ref{eq:ev_set_adS}) is, as in Minkowski space-time,
divergent as the coincidence limit is taken. We follow our earlier procedure
in Minkowski space-time, subtracting the vacuum Feynman Green's function
$S^{\mathrm {adS}}_{\mathrm {vac}}$ from the thermal Feynman Green's function $S^{\mathrm {adS}}_{\beta }$,
and then substituting into (\ref{eq:ev_set_adS}). This gives a finite quantity as $x'\rightarrow x$.
We thus calculate the difference between the t.e.v.~and the v.e.v.~of the SET, which we denote by
$\braket{:T_{\hat{\alpha}\hat{\rho}}:}^{\rm adS}_{\beta  }$.
The details are presented in \cite{Ambrus:2017cow}, where we find that
$\braket{:T_{\hat{\alpha}\hat{\rho}}:}^{\rm adS}_{\beta  }$ has the perfect fluid form
(\ref{eq:fluid}), with energy density $E^{\mathrm {adS}}_{\mathrm {QFT}}(\beta )$
 given in the massless limit $m\rightarrow 0$ by
\begin{equation}
E^{\mathrm {adS}}_{\mathrm {QFT}}(\beta ) = \frac{3\omega ^4}{4\pi^2} \cos ^{4}\omega r   \sum_{j = 1}^\infty (-1)^{j-1}
\frac{\cosh\frac{\omega j \beta}{2}}{\sinh ^{4}\frac{\omega j \beta}{2} },
\label{eq:qft_m0}
\end{equation}
while the pressure is $P^{\mathrm {adS}}_{\mathrm {QFT}}(\beta ) = E^{\mathrm {adS}}_{\mathrm {QFT}}(\beta ) / 3$.
The results for massive fermions are discussed in detail in~\cite{Ambrus:2017cow}.

As in the RKT calculation, $E^{\mathrm {adS}}_{\mathrm {QFT}}(\beta )$ and
$P^{\mathrm {adS}}_{\mathrm {QFT}}(\beta )$ (\ref{eq:qft_m0}) are
not constant on adS space-time, but depend on the radial coordinate $r$.
By studying their profiles at different values of $\beta $, it is found in
\cite{Ambrus:2017cow} that quantum corrections are more significant at
larger values of $\beta $, corresponding to lower temperatures. For small
$\beta $ (and high temperature), quantum corrections are less significant
and the fermions behave essentially classically.

\begin{figure}
\begin{tabular}{cc}
 \includegraphics[width=.45\linewidth]{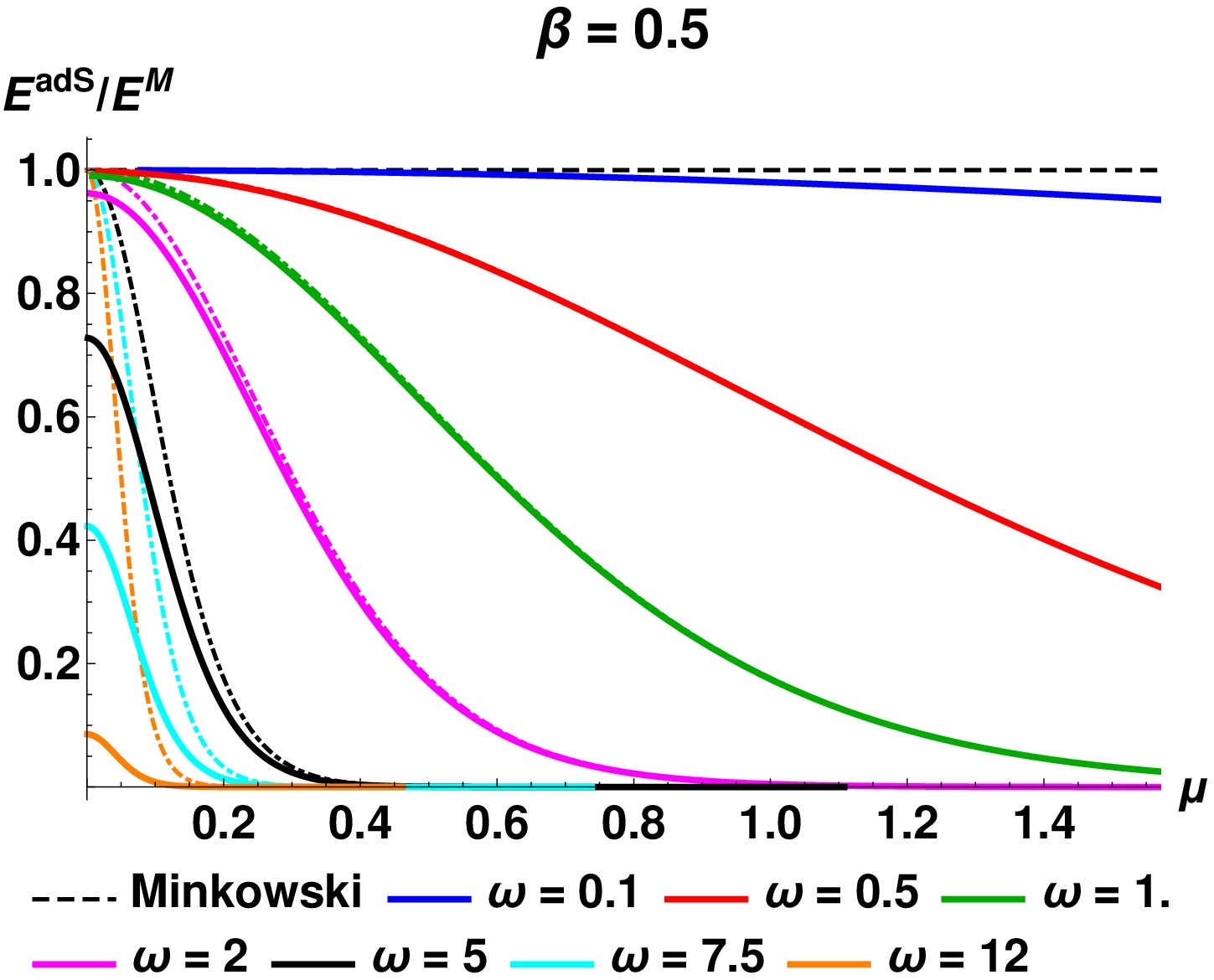} & \quad
 \includegraphics[width=.45\linewidth]{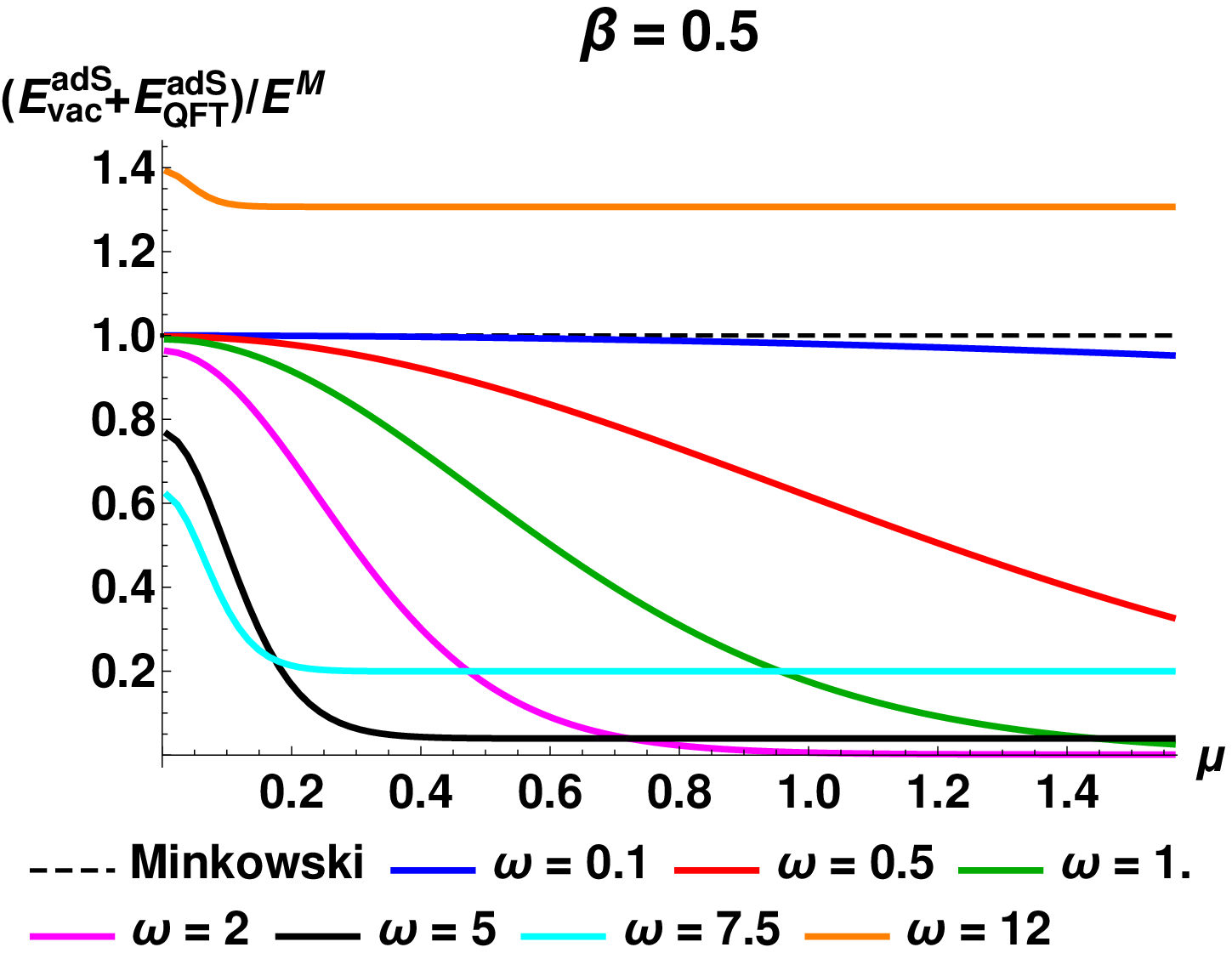}\\
(a) & (b)
\end{tabular}
\caption{(a) $E^{{\mathrm {adS}}}_{\mathrm {QFT}}(\beta )$ \eqref{eq:qft_m0} (solid lines)
and $E^{{\mathrm {adS}}}_{\mathrm{RKT}}(\beta )$ \eqref{eq:mzeroadS} (dotted lines).
(b) The sum of $E^{{\mathrm {adS}}}_{\mathrm {QFT}}(\beta )$ \eqref{eq:qft_m0} and the vacuum energy density
$E_{\mathrm {vac}}^{\mathrm {adS}}$ \eqref{eq:vac_m0}.
The results are for massless particles at $\beta = 0.5$ and are
normalized with respect to $E^{\mathrm{M}}(\beta)$ (\ref{eq:mzeroM}).}
\label{fig:profiles}
\end{figure}

The analysis in \cite{Ambrus:2017cow} is performed at constant inverse radius of curvature $\omega$.
In Fig.~\ref{fig:profiles} (a) we consider instead
the energy
densities $E^{\mathrm {adS}}_{\mathrm {RKT}}(\beta )$ (\ref{eq:mzeroadS}) and
$E^{\mathrm {adS}}_{\mathrm {QFT}}(\beta )$ (\ref{eq:qft_m0})
divided by the corresponding Minkowski energy
$E^{\mathrm{M}}(\beta ) $ (\ref{eq:mzeroM})
when the inverse temperature $\beta $
is fixed and the inverse radius of curvature $\omega$ varies. For simplicity, we only
consider the results for massless particles.
Since the adS boundary is an infinite proper distance from the origin, in Fig.~\ref{fig:profiles} we show how
the energy
densities depend on
the geodetic distance $\mu _{\mathrm {adS}}$ between the origin and a point at radial coordinate $r$ in adS, given by
\begin{equation}
 \mu _{\mathrm {adS}}(r) = \int_{0}^r ds = \omega ^{-1} {\mathrm {arccosh}} \left( \sec \omega r\right),
 \label{eq:mu}
\end{equation}
where \eqref{eq:metric} was used for the adS line element $ds$.
On Minkowski space-time, the corresponding quantity is $\mu _{\mathrm {M}}(r)=r$.
Two distinctive features can be observed
in Fig.~\ref{fig:profiles}. First, the factor of $\cos ^{4} \omega r$ in (\ref{eq:qft_m0}) causes the local temperature ${\tilde {\beta }}^{-1}$ (\ref{eq:tolman}), and thus the energy density,
to decrease at large values of $\mu _{\mathrm {adS}}$ (corresponding to $\omega r \rightarrow \pi /2$). This feature is also predicted by RKT \eqref{eq:mzeroadS}.
The second important feature is that the value of $E^{\mathrm{adS}}_{\mathrm{QFT}}(\beta)$
at the origin $r = 0$ ($\mu _{\rm {adS}}=0$) decreases as $\omega$ increases. This is in contrast to the RKT result
\eqref{eq:mzeroadS}, which predicts that $E^{\mathrm{adS}}_{\mathrm{RKT}}(\beta)$ is independent
of $\omega$ at $r =0$, its value being given by the Minkowski expression \eqref{eq:mzeroM}.
As concluded in \cite{Ambrus:2017cow}, quantum corrections quench the thermal energy density, and this effect becomes more pronounced as the adS curvature increases.

\section{Vacuum contribution to thermal expectation values}

The t.e.v.~of the SET $\braket{T_{\hat{\alpha}\hat{\rho}}}_\beta^{\mathrm {adS}}$  on adS space-time can be written in terms of the v.e.v.~$\braket{T_{\hat{\alpha}\hat{\rho}}}_{\rm vac}^{\mathrm {adS}}$ as
\begin{equation}
 \braket{T_{\hat{\alpha}\hat{\rho}}}_\beta^{\mathrm {adS}} = \braket{T_{\hat{\alpha}\hat{\rho}}}_{\mathrm {vac}}^{\mathrm {adS}} +
 \braket{:T_{\hat{\alpha}\hat{\rho}}:}^{\rm adS}_\beta .
\label{eq:exp}
\end{equation}
The renormalized v.e.v.~of the SET $\braket{T_{\hat{\alpha}\hat{\rho}}}_{\mathrm {vac}}^{\mathrm {adS}}$
can be written in perfect fluid form $\braket{T_{\hat{\alpha}\hat{\rho}}}_{\mathrm {vac}}^{\mathrm {adS}} = {\mathrm {Diag}}
\{ E_{\mathrm {vac}}^{\mathrm {adS}}, P_{\mathrm {vac}}^{\mathrm {adS}}, P_{\mathrm {vac}}^{\mathrm {adS}}, P_{\mathrm {vac}}^{\mathrm {adS}} \} $.
For nonzero fermion mass, $\braket{T_{\hat{\alpha}\hat{\rho}}}_{\mathrm {vac}}^{\mathrm {adS}}$ is not uniquely defined, since it
depends on the renormalization method and on an arbitrary renormalization mass scale \cite{Ambrus:2015mfa}.
However, in the massless case,
$E_{\mathrm {vac}}^{\mathrm {adS}}$ and $P_{\mathrm {vac}}^{\mathrm {adS}}$ are independent of the method of renormalization and the arbitrary mass scale and are given by \cite{Ambrus:2015mfa}:
\begin{equation}
 E_{\mathrm {vac}}^{\mathrm {adS}} = -P_{\mathrm {vac}}^{\mathrm {adS}} = \frac{11\omega^4}{960 \pi^2}.\label{eq:vac_m0}
\end{equation}
 The v.e.v.~(\ref{eq:vac_m0}) is constant throughout adS space-time and vanishes in the Minkowski limit $\omega \rightarrow 0$.
The fact that $P_{\mathrm {vac}}^{\mathrm {adS}}\neq E_{\mathrm {vac}}^{\mathrm {adS}}/3$ is due to the usual conformal anomaly
in curved space-time.

Figure \ref{fig:profiles} (b) shows $\braket{T_{\hat{\alpha}\hat{\rho}}}_\beta^{\mathrm {adS}}$ (\ref{eq:exp}) as a function of $\mu _{\mathrm {adS}}$ (\ref{eq:mu}) for fixed inverse temperature $\beta $.
The contribution of the v.e.v.~$\braket{T_{\hat{\alpha}\hat{\rho}}}_{\mathrm {vac}}^{\mathrm {adS}}$ can be clearly seen by comparing Fig.~\ref{fig:profiles} (a) and Fig.~\ref{fig:profiles} (b).
For a fixed value of $\beta$, there
is always a vicinity of the space-time boundary where the vacuum contribution will dominate, since
\eqref{eq:qft_m0} indicates that the t.e.v. of the SET vanishes near the boundary as
$\cos ^{4}\omega r $, while the v.e.v.~retains its constant value \eqref{eq:vac_m0}.

\begin{figure}
\includegraphics[width=.45\linewidth]{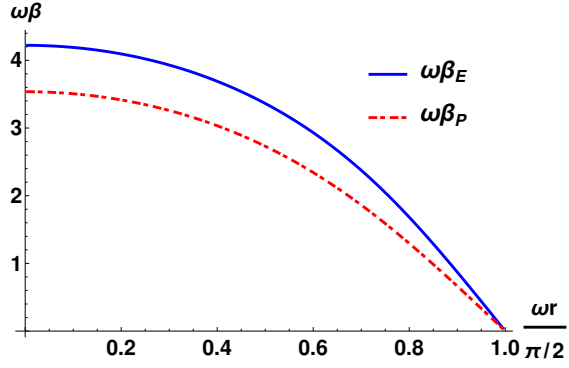}
\caption{The critical inverse temperatures $\omega \beta_E$
and $\omega \beta_P$
\eqref{eq:thvsvac} as functions of $\omega r$.}
\label{fig:thvsvac}
\end{figure}

For fixed radius $r < \pi / 2\omega$, and sufficiently large $\beta $ (small temperature), $\braket{:T_{\hat{\alpha}\hat{\rho}}:}_{\rm adS}^\beta$ is negligible compared to the v.e.v., while for sufficiently small $\beta $ (large temperature), the v.e.v.~is negligible.
There is therefore a critical inverse temperature $\beta_E$ at which the thermal
energy density $E^{\mathrm {adS}}_{\mathrm {QFT}}(\beta_E)$ (\ref{eq:qft_m0}) equals the v.e.v.~of the energy density $E_{\mathrm {vac}}^{\mathrm {adS}}$ (\ref{eq:vac_m0}).
Similarly, there is a critical inverse temperature $\beta_P$ at which
the pressure obeys $P^{\mathrm {adS}}_{\mathrm {QFT}}(\beta_P) =-P_{\mathrm {vac}}^{\mathrm {adS}}$.
The critical inverse temperatures $\beta_E$ and $\beta _{P}$ are solutions of the equations:
\begin{equation}
 \frac{3\omega ^4}{4\pi^2} \cos ^{4}\omega r   \sum_{j = 1}^\infty (-1)^{j-1}
\frac{\cosh(\omega j \beta_E/2)}{\sinh ^{4}(\omega j \beta_E/2) } =
 \frac{11\omega^4}{960 \pi^2}, \qquad
 \frac{\omega ^4}{4\pi^2} \cos ^{4}\omega r   \sum_{j = 1}^\infty (-1)^{j-1}
\frac{\cosh(\omega j \beta_P/2)}{\sinh ^{4}(\omega j \beta_P/2) } =
 \frac{11\omega^4}{960 \pi^2},
 \label{eq:thvsvac}
\end{equation}
and are shown in Fig.~\ref{fig:thvsvac} as functions of $\omega r$.
Both $\omega \beta_E$ and $\omega \beta_P$ decrease monotonically as $\omega r$ increases and vanish on the boundary, where the thermal energy density and pressure are zero.
The critical inverse temperature $\beta _{E}$ is larger than $\beta _{P}$, indicating that thermal effects dominate in the energy density at a lower temperature than for the pressure.  

\section{CONCLUSIONS}
\label{sec:conc}

In this paper we have considered the vacuum contributions to the t.e.v.~of the
SET on adS space-time. We first showed that on Minkowski space-time,
the quantum SET at finite inverse temperature $\beta$ is exactly equal to the one predicted by RKT.
On adS, we have investigated the quenching effect of quantum corrections on the
t.e.v.~of the SET, taken with respect to the vacuum state, by
increasing the inverse radius of curvature $\omega$. Finally, we have shown that the full
t.e.v.~of the SET, comprising the vacuum and thermal contributions, exhibits a non-monotonic behaviour as $\omega $ increases,
since the vacuum contribution becomes dominant at large values of $\omega$.

\section{ACKNOWLEDGMENTS}
V.E.A.~was supported by a grant from the
Romanian National Authority for Scientific Research and Innovation,
CNCS-UEFISCDI, project number PN-II-RU-TE-2014-4-2910.
The work of E.W.~is supported by the Lancaster-Manchester-Sheffield Consortium for
Fundamental Physics under STFC grant ST/L000520/1.

\end{document}